\documentclass[lettersize,journal]{IEEEtran}
\usepackage{amsmath,amssymb,amsfonts,bm} 
\usepackage{algorithmic}
\usepackage{algorithm}
\usepackage{array}
\usepackage{caption}
\usepackage{subcaption}
\usepackage{textcomp}
\usepackage{stfloats}
\usepackage{bigints}
\usepackage{url}
\usepackage{verbatim}
\usepackage{graphicx}
\usepackage{mathtools}
\usepackage{dsfont}
\usepackage{esvect}
\usepackage{accents}
\usepackage{cite}
\usepackage{pbalance}
\usepackage{xcolor}
\newcolumntype{M}[1]{>{\centering\arraybackslash}m{#1}}
\newcommand*{\QEDA}{\null\nobreak\hfill\ensuremath{\blacksquare}}%

\newcommand\myeqa{\mathrel{\stackrel{\makebox[0pt]{\mbox{\normalfont\scriptsize (a)}}}{=}}}
\newcommand\myeqb{\mathrel{\stackrel{\makebox[0pt]{\mbox{\normalfont\scriptsize (b)}}}{=}}}
\newcommand\myeqc{\mathrel{\stackrel{\makebox[0pt]{\mbox{\normalfont\scriptsize (c)}}}{=}}}

\newcommand\myeqdef{\mathrel{\stackrel{\makebox[0pt]{\mbox{\normalfont\scriptsize conv}}}{=}}}

\begin{document}

\title{Performance Analysis and Experimental Validation of UAV Corridor-Assisted Networks}
%
%
%

\author{Harris K.~Armeniakos,~\IEEEmembership{Member,~IEEE,}
         Viktor~Nikolaidis,~\IEEEmembership{Member,~IEEE,}
        Petros S.~Bithas,~\IEEEmembership{Senior Member,~IEEE,}
         Konstantinos Maliatsos,~\IEEEmembership{Member,~IEEE,}
       and Athanasios G.~Kanatas,~\IEEEmembership{Senior Member,~IEEE}
\thanks{H. K. Armeniakos, V. Nikolaidis and A. G. Kanatas are with the University of Piraeus, Piraeus, Greece (e-mail: \{harmen, vnikola, kanatas\}@unipi.gr). P. S. Bithas is with the National and Kapodistrian University of Athens, Athens, Greece (e-mail:pbithas@dind.uoa.gr). K. Maliatsos is with the University of the Aegean, Samos, Greece (e-mail: kmaliat@aegean.gr).}
\thanks{The publication of this paper has been partly supported by the University of Piraeus Research Center. iSEE-6G has received funding from the Smart Networks and Services Joint Undertaking (SNS JU) under the European Union’s Horizon Europe research and innovation programme under Grant Agreement No 101139291.}
}
\maketitle

\begin{abstract}
Unmanned aerial vehicle (UAV) corridor-assisted communication networks are expected to expand significantly in the upcoming years driven by several technological, regulatory, and societal trends. In this new type of networks, accurate and realistic channel models are essential for designing reliable, efficient, and secure communication systems. In this paper, an analytical framework is presented that is based on one-dimensional (1D) finite point processes, namely the binomial point process (BPP) and the finite homogeneous Poisson point process (HPPP), to model the spatial locations of UAV-Base Stations (UAV-BSs). To this end, the shadowing  conditions experienced in the UAV-BS-to-ground users links are accurately considered in a realistic maximum power-based user association policy. Subsequently, coverage probability analysis under the two spatial models is conducted, and exact-form expressions are derived. In an attempt to reduce the analytical complexity of the derived expressions, a dominant interferer-based approach is also investigated. Finally, the main outcomes of this paper are extensively validated by empirical data collected in an air-to-ground measurement campaign. To the best of the authors’ knowledge, this is the first work to experimentally verify a generic spatial model by jointly considering the random spatial and shadowing characteristics of a UAV-assisted air-to-ground network.
 
\end{abstract}


%
\IEEEpeerreviewmaketitle

\section{Introduction}
\IEEEPARstart{A}{s we} step toward the sixth-generation (6G) communication networks, unmanned aerial vehicles (UAVs) are envisioned to be a core pillar of the Internet-of-Things (IoT) networks that aim to realize massive connections between several devices \cite{IoT}. In various use cases, UAVs are considered to cooperatively operate in swarms to serve ground devices for delivery services \cite{alqudsi2025uav}, communication or energy transfer \cite{swarms}. As the number of UAVs increases, UAV traffic management (UTM) has gained an important interest from various authorities such as the Federal Aviation Administration (FAA) and the National Aeronautics and Space Administration (NASA) \cite{FAA}. In particular, similar to ground vehicles and piloted aircraft, as UAVs become more common, their movement might be limited to specific aerial highways, also known as UAV corridors, which should also comply with air traffic regulation authorities \cite{cherif20213d,karimi2024optimizing}. The main objectives of these air corridors are to minimize collision risks, avoid disruption to existing air traffic, and facilitate integration with the current airspace management system. Mandating UAVs to fly within these corridors, operators shift the focus from providing network services throughout the sky to ensuring reliable connectivity within these corridors. For example, in the United States, the FAA sets the parameters for air corridors within class B, C, or D airspaces \cite{namuduri2022advanced}. The development of air corridors is still in an early stage, as various aspects must be decided such as traffic regulations, safety standards, and performance criteria. 

Various recent contributions can be found in the technical literature for air-corridor communications systems. For example, in \cite{karimi2024optimizing}, the necessary conditions are identified and iterative algorithms are designed to optimize antenna tilts and transmit power at each base station (BS) of a cellular network to provide the best quality of service to both ground users and UAVs flying along vertical line corridors.  In \cite{swarms}, assuming linear UAV corridor networks, the strongest BS signal association policy is considered, which provides better coverage than simply connecting to the nearest BS.  In \cite{10569086}, assuming that the UAVs are located on a fixed height aerial highway, a novel approach is proposed to take advantage of the knowledge of the aerial highway to optimally plan the fifth generation (5G) synchronization signal block beams. 
In \cite{ellis2023interference}, assuming a communication network consisting of numerous equal distance parallel drone corridors, the interference of UAV-to-UAV communications is investigated, while an interference mitigation strategy is also proposed for millimeter wave (mmWave) beamforming. Finally, in \cite{zaid2025aerial}, in a heterogeneous two-tier communication network, where air corridors have been modeled using the Poisson line process, the probability of coverage has been analytically investigated using tools from stochastic geometry.  Based on the referenced bibliography, it is evident that regardless of the aerial corridor topology (e.g., line, lane, star) or the type of research (e.g., performance analysis), UAV corridor-assisted networks are anticipated to attract significant interest from both academic and standardization sectors. Fortunately, many impactful UAV use cases could still be enabled by providing reliable connectivity along predetermined aerial routes, i.e., UAV corridors, enforced by the appropriate traffic authorities and have been reported in detail in \cite{cherif20213d}. In fact, some key application areas include i) public safety, ii) post-disaster relief and emergency communications, and iii) extension of coverage in rural areas.


Over the last years, stochastic geometry has emerged as a very powerful tool for modeling and analyzing the performance of complicated UAV-assisted networks, revealing key design insights. When a fixed and predetermined number of UAVs are deployed in a finite region to provide coverage or to execute a certain task, their spatial distribution can be well described by the simple but reasonable homogeneous binomial point process (BPP). The BPP spatial modeling of UAVs has been widely adopted for the performance analysis of finite UAV or UAV-assisted  networks \cite{Dhillon, Leung, Krishna, 9740446}.  In \cite{frontiers}, for the first time a stochastic geometry framework assisted by UAV corridors was proposed.  However, in real-world application scenarios, the number of UAVs deployed in a finite region may be random or not always predetermined. Moreover, some of the UAVs deployed in certain regions are usually in idle or standby mode to maintain the network's energy efficiency,  making also the number of active UAVs random. In such cases, the finite homogeneous Poisson point process (HPPP) is a more suitable model compared to the BPP spatial model, as different realizations of the process consist of a different number of points. Consequently, in \cite{blockages}, the authors conduct a performance comparison in terms of coverage probability between the finite HPPP and the BPP spatial modeling of UAVs. However, as more UAVs are deployed, both spatial models tend to exhibit a similar behavior \cite{blockages}. From a clearly mathematical point of view, the analytical framework when BPP is used is more challenging than that with HPPP \cite{Dhillon}. For example, if the reference receiver is served by the serving UAV-BS, the distribution of the BPP (after the serving UAV-BS is removed) is not the same as that of the original BPP. Moreover, the performance is location dependent. This means that conditioned on the serving BS, the derivation of the conditional distances between the receiver and the interfering BSs is required in the statistical characterization of the Laplace transform of aggregate interference. Moreover, BPP is more suitable when a fixed number of UAVs are deployed to cover a small-scale area, while finite HPPP is preferable for large-scale deployments \cite{blockages}.  In fact, the HPPP serves as a fundamental point process for modeling and evaluating the performance of large-scale UAV-assisted networks mainly due to its analytical tractability. However, both the HPPP and the BPP can accurately capture the spatial randomness and random deviations of the UAVs from a well-defined spatial arrangement. Note that a key challenge in performance analysis when employing the finite HPPP is that one has to consider the probability that the finite set be empty.

In UAV-to-ground communications, the wireless medium is usually characterized by line-of-sight (LoS) propagation conditions as a result of the UAVs' ability to hover above the buildings. However, in several cases, the LoS assumption is not satisfied, due to the presence of large obstacles. In such scenarios, large-scale fading (shadowing) occurs, resulting in randomly varying envelope mean levels. The impact of shadowing on the performance of UAV-assisted networks has been extensively investigated through stochastic analysis \cite{Bithas,Bithas2,shadowing,shadowing2}. In fact, shadow effects are modeled using distributions that fit well with empirical data \cite{Proceedings}. However, this approach has not been adopted in prior stochastic geometry-based frameworks, in which the existence of blockages has been modeled using a probability distribution function for the LoS conditions, allowing for a binary-valued path loss factor \cite{blockages}, \cite{shadowingPL,shadowingPL2,shadowingPL3}. Note that a few works \cite{Direnzo1}, \cite{Direnzo2} have captured the effect of large-scale fading on the performance of terrestrial stochastic geometry-based cellular networks. To the best of the authors' knowledge, the adoption of an empirical distribution for modeling shadowing effects in UAV stochastic geometry-based frameworks is a new research approach that should be carefully investigated.

 Among the numerous applications of UAVs, UAV-enabled airborne communication has recently attracted notable research attention. In this paradigm, dedicated UAVs are used as aerial BSs, access points, or relays to assist wireless communications of ground nodes, which we refer to as UAV-assisted wireless communications. In UAV-assisted networks, terrestrial user equipment (UE) can be attached to UAV-BS to establish communication. The user association policy to determine the serving UAV-BS is based on the association criterion for maximum average received power, and the receiver is agnostic to the conditions that provide maximum received power. If the impact of shadowing is ignored in the calculation of the maximum average received power, then the latter is reduced to the conventional minimum Euclidean distance association criterion, as shown in \cite{shadowingPL4}. However, in this work, it is rigorously argued that if a realistic modeling of the random fluctuations of the average received signal power is required, then the impact of shadowing in the maximum received power association policy should be considered.

Motivated by the aforementioned, in this work, the performance of a UAV corridor-assisted communication system in the presence of shadowing is analytically investigated based on the stochastic geometry tool. To this end, two cases for modeling the spatial locations of UAV-BSs in a line corridor\footnote{The closest geometric shape to approximate a UAV corridor is the cylinder. Unfortunately, conducting a performance analysis in a UAV corridor-assisted network, whose UAV corridor is modeled as a cylinder, will result in extremely high analytical and computational complexity even if interference is neglected \cite{9740446}. To balance the trade-off between accuracy and analytical complexity, the UAV corridor is modeled as a finite line segment above the ground, which is a simple yet reasonable geometric model to capture the concept of a UAV corridor.} above the ground are considered, namely the one-dimensional (1D) BPP and the 1D finite HPPP. By accurately modeling shadowing as an inverse gamma (IG) distributed random variable (RV)\footnote{IG distribution has been proposed as a shadowing model, since it offers an excellent fit to measurement data for large-scale fading, while maintaining analytical tractability \cite{Bithas}.}, a maximum average received power user association policy is proposed that \emph{jointly} considers both path loss and shadowing conditions of UAV-BSs, and a comprehensive coverage probability analysis is conducted.  It should be mentioned that the performance of a UAV corridor-assisted network is more challenging than that of its two-dimensional (2D) counterpart in a disc above ground. This occurs because adopting a maximum power-based user association policy, both the derivation of the probability density function (PDF) of the maximum received power and the statistical characterization of the Laplace transform of the aggregate interference cannot be obtained in closed form. The latter is due to the fact that the expressions of the relevant distance distributions contain square roots. To simplify the analytical results, a dominant interferer approach is also presented. Finally, theoretical derivations are compared with empirical results, obtained from measurement data, which demonstrate the impact of large-scale fading on the performance of UAV-to-ground communications.  

More specifically, the main contributions of this paper are as follows. 

\subsubsection{Refining the maximum power association policy in UAV corridor-assisted networks} In realistic UAV-assisted networks in urban environments, the selection of the serving UAV-BS depends both on its random spatial location and on the shadowing conditions. As a key contribution,  a realistic maximum power user association policy is considered that \emph{jointly} uses both Euclidean distances and the shadowing conditions of the UAV-BSs in the calculation of the maximum received power. 

\subsubsection{Performance analysis under 1D BPP spatial modeling of UAV-BSs in an aerial corridor} By employing the 1D BPP to model the spatial locations of UAV-BSs, a fixed and predetermined number of UAV-BSs are deployed in an aerial corridor modeled as a line segment above the ground. Subsequently, by adopting the maximum power association policy for selecting the serving UAV-BS, performance analysis is conducted in terms of coverage probability. As a key intermediate result, the PDF of the maximum received power is derived in exact form. Next, in an attempt to simplify the analytical findings, a dominant interferer scenario is also analytically investigated. 

\subsubsection{ Performance analysis under 1D finite HPPP spatial modeling of UAV-BSs in an aerial corridor} By employing, for the first time the 1D finite HPPP to model the spatial locations of the UAV-BSs, a coverage probability analysis is performed under aggregate interference. Then, the Laplace transform of the aggregate interference power distribution accounting for both the shadowing conditions and the locations of the interfering UAV-BSs is calculated.

\subsubsection{Experimental validation of the analytical results using measured data} As a key contribution, the performance in terms of coverage probability for both spatial models is validated by empirical data obtained from a measurement campaign to verify the proposed framework and analysis. Moreover, the maximum received power association policy is compared both empirically and through simulations to a conventional one, assuming the minimum Euclidean distance criterion for selecting the serving UAV-BS.  As a step further, an analysis for a variable height model of the UAV-BSs is conducted and experimentally validated. Consequently, the Normal distribution is shown to be the most reasonable distribution for capturing the random height variations of the UAV-BS. Finally, several meaningful key insights are drawn for the design and performance of UAV corridor-assisted networks in urban environments. 

The remainder of this paper is organized as follows. In Section II, the system model of the aerial corridor network is provided, in conjunction with the corresponding spatial and channel models. In Section III, an analytical mathematical framework for the coverage probability is presented using BPP spatial modeling, while in Section IV, the coverage probability analysis is based on HPPP spatial modeling. In Section V, various numerically evaluated, simulated, and experimental results are presented and compared. Finally, the conclusions of this article are drawn in Section VI.

\section{System and Channel Models}
\subsection{Network Model}
Consider a downlink finite UAV-assisted network that integrates terrestrial UEs and UAV-BSs. The UEs are assumed to be uniformly and independently distributed according to a stationary point process in a finite area $\mathcal{A} \subset \mathbb{R}^2$. For this setup, it is assumed that $\mathcal{A}$ is a two-dimensional (2D) circular area $b(\mathbf{o},R)$ centered at the origin $\mathbf{o} = [0,0]^{T}$ with radius $R$, i.e., $\mathcal{A}=b(\mathbf{o},R)$ as shown in Fig. 1. Consider that the UAV-BSs hover in a UAV aerial corridor according to some uniform point process, as will be discussed in the next subsections, with rules enforced by appropriate traffic authorities. The UAV corridor is modeled as a line segment, located $h$ meters above the ground. Without loss of generality, the on-ground projection of the center of the UAV corridor, namely $\mathbf{o^{'}}=[0,0,h]^{T}$, crosses the origin of the coordinates $\mathbf{o}$. Consequently, it is assumed that the UAV-BSs are uniformly and independently distributed on the line segment $L(-R,R)$, with $L(-R,R)$ denoting a finite line segment centered at $\mathbf{o^{'}}$ extending from $-R$ to $R$ with length $|L(-R,R)| = 2 R$. Accordingly, the spatial locations of UAV-BSs form a 1D uniform point process. Moreover, it is assumed that the receiving device is located at the ground origin $\mathbf{o}$ at a given time. After averaging the performance of this receiving UE over the corresponding point process  BPP or finite HPPP, it becomes the typical receiving UE which will
be interchangeably referred to as just \emph{receiver}.  Note that the proposed framework does not prohibit multiple ground users from being deployed in the terrestrial region and each one to be served by a single UAV-BS. A representative illustration of the aforementioned model is shown in Fig. 1. In the following, the two uniform point processes that are used to model the spatial locations of UAV-BSs are presented.

\subsubsection{BPP Spatial Modeling of UAV-BSs} In this setup, a predefined and fixed number of $N$ UAV-BSs is assumed to be uniformly and independently distributed on $L(-R,R)$ and their spatial locations form a 1D BPP $\Psi$.  
Subsequently, let $\{\mathbf{y}_i\} \equiv \Psi$ denote the spatial locations of the UAV-BSs. Then, let $r_i = \| \mathbf{y}_i - \mathbf{o^{'}} \|$, for $1 \leq i \leq N$, denote the Euclidean distance of the $i$-th UAV-BS from $\mathbf{o^{'}}$. Accordingly, $\{d_i\}_{i=1:N} = \{\sqrt{r_i^2 + h^2}\}$ denotes the unordered set of Euclidean distances between the receiver and the UAV-BSs. 

\subsubsection{Finite HPPP Spatial Modeling of UAV-BSs} In an alternative setup, the locations of UAV-BSs hovering in a UAV aerial corridor form an almost surely (a.s.) 1D finite HPPP $\Phi$ with intensity $\lambda_{UAV}$, defined as $\Phi=\mathcal{W}\cap L(-R,R)$, where $\mathcal{W}$ denotes the 1D HPPP of intensity $\lambda_{UAV}$. For a UAV-BS located at $y \in \Phi$, let $r_y = \|y-\mathbf{o^{'}} \|$ and $d_y$ denote the Euclidean distance between the receiver and a UAV-BS $y$ as $d_y =\sqrt{r_y^2 + h^2}$.

\subsection{Path Loss and Channel Models}

\subsubsection{Path loss}
The average channel conditions are characterized by different path loss exponents, denoted by $\alpha$. Then, following the standard power-law path-loss model for the link between the receiver and a UAV-BS located at $\mathbf{y}_i \in \Psi$ or $y \in \Phi$, the random path loss function is defined as
\begin{equation}
l(d_k) = d_k^{-\alpha},  
\end{equation}
where $k \in\{i, y\}$. By further considering the carrier frequency $f_c$ through the factor $K = \big(\frac{c}{4 \pi f_c}\big)^2$, the path loss function is rewritten as $l^{'}(d_k) = K\, l(d_k)$, where $c$ denotes the speed of light.

\begin{figure}[!t]
    \centering
    \includegraphics[keepaspectratio,width=3.45in]{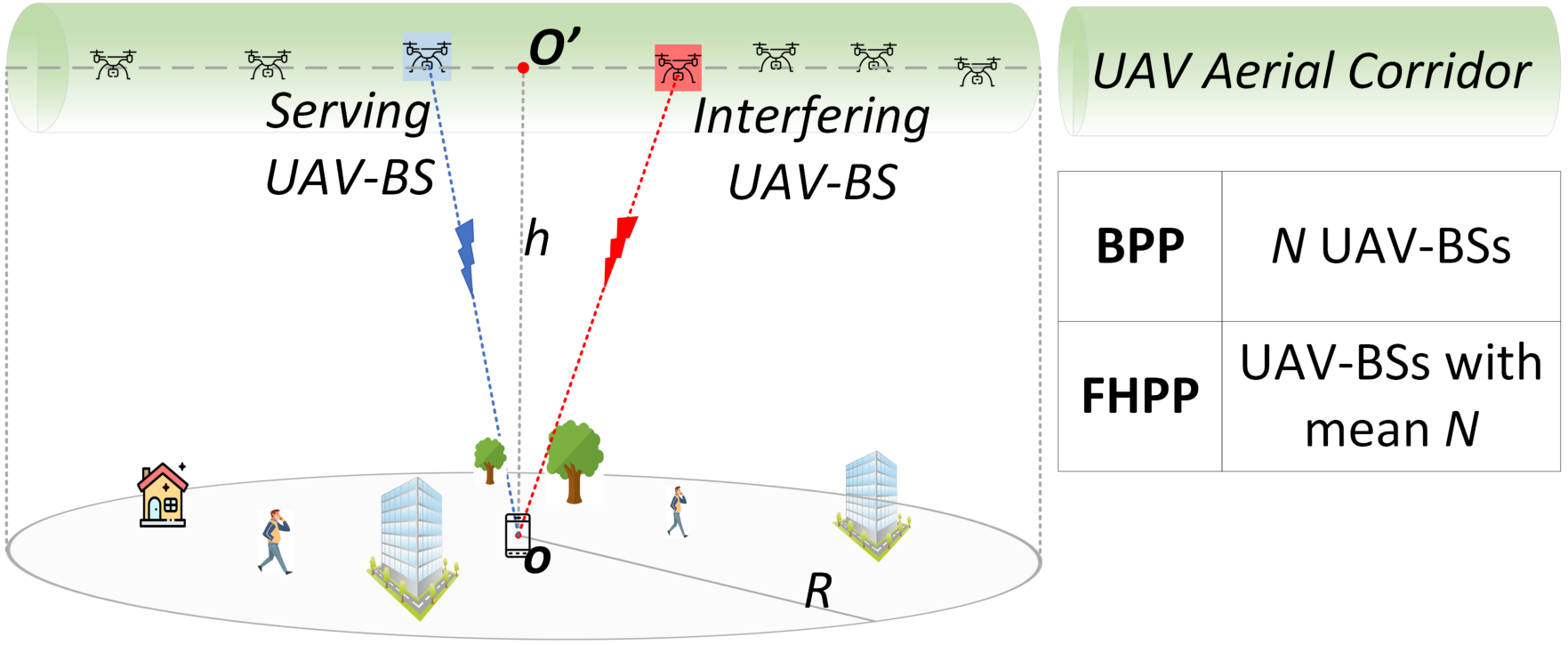}
       \caption{ Illustration of the UAV corridor-assisted network. The serving UAV-BS is chosen by considering the shadowing conditions of UAV-BSs in the user association policy.}
    \label{fig:Fig2}
\end{figure}

\subsubsection{Large-scale fading}
In general, the radio signals emitted by the UAV transmitter propagate in free space until reaching the urban environment where they undergo shadowing and scattering caused by man-made structures, forested or vegetated areas, and mountainous or Hilly Terrain. This line-of-sight blockage is directly related to the presence of shadowing, i.e., random variations of the mean envelope level, since an increase in non-LoS (NLoS) probability results in severe shadowing \cite{stuber2001principles}. The random fluctuations of these local means can be modeled with RV $S_{k}$, with $k \in\{i, y\}$. In aerial-to-ground communication scenarios, a widely adopted distribution that offers accurate approximation to shadowing coefficients and mathematical tractability is the IG distribution \cite{Bithas}. The PDF of the IG distribution is given by
 \begin{equation}
f_{S_k}(x) =  \frac{\gamma^q}{\Gamma(q) x^{q+1}} \exp\Big(-\frac{\gamma}{x}\Big), 
\end{equation}
where  $q > 1$ is the shaping parameter of the distribution, related to the severity of the shadowing, i.e., increased values of $q$ result in higher NLoS probability and $\gamma$ denotes the scaling parameter. Moreover, $\Gamma(\cdot)$ is the Gamma function \cite[eq. (8.310.1)]{Ryzhik}.

\subsubsection{Small-scale fading}
Fading effects are also taken into account, and the signal amplitude is assumed to follow the Nakagami-$m$ distribution.  By adopting the Nakagami-$m$ for modeling the amplitudes of the small-scale fading channel, the channel power gains $h_{k}$, with $k \in\{i, y\}$, following the Gamma distribution, that is, $h_k \sim $ Gamma$\big(m,\frac{1}{m}\big)$ with the shape and scale parameters of $h_k$ being $m$ and $1/m$, respectively. Note that $\mathbb{E}[h_k]=m \frac{1}{m}=1$. The PDF of $h_k$ is given by
\begin{equation}
  f_{h_k}(x) = \frac{{m}^{m} x^{m-1}}{\Gamma(m)}\exp{(-m x)}.
\end{equation}

\section{BPP Performance Analysis}
\subsection{User Association Policy and SIR Definition}

\subsubsection{User Association Policy}\label{sub:user_association} In wireless networks, the receiver is agnostic to the conditions that provide maximum power and selects the serving BS, within a distance-limited finite area, among a set of candidate serving BSs. Note that triggered by the 5G New Radio (NR) beam management procedure, the receiver selects the serving BS by measuring the received power of all UAV-BS.

According to this principle, by considering a maximum power user association policy, the receiver is associated with the UAV-BS providing the maximum average received power within $\Psi$. In this work, both distance-dependent path loss and large-scale fading are considered \emph{jointly} in the calculation of the maximum received power. Accordingly, the maximum received power is given by 
\begin{equation}
Pr_{0} = \underset{\begin{subarray}{c}
  i = 1:N
  \end{subarray}}{\operatorname{max}}\{Pr_i\} \\ 
  = \underset{i = 1:N} {\operatorname{max}}\{{S_i}\, l(d_i)\}.   
\end{equation}
Notably, in this policy, the serving UAV-BS is not necessarily the closest UAV-BS in Euclidean distance.

\subsubsection{SIR Definition}
In the general case, where all UAV-BSs share the same resource blocks\footnote{In this work, a frequency reuse factor equal to 1 is considered, i.e. all UAV-BSs use the same frequency resources. It should be noted that performance evaluation of UAV-assisted communications using the same frequency channel is extremely common in the open technical literature. This is because a frequency reuse factor equal to 1 corresponds to a worst-case interference scenario and therefore can reveal meaningful information about network performance under aggregate interference}., then $(N-1)$ of them interfere with the receiver. The received signal-to-interference ratio (SIR)\footnote{Under aggregate interference, the network tends to be interference-limited and therefore the noise power is assumed to be negligible as compared to the aggregate interference power \cite{interference}.} is given by 
\begin{equation}
{\rm{SIR}} = \frac{h_{0} Pr_{0}}{I}, 
\end{equation}
where $I$ refers to the aggregate interference power and is given by $I = \sum_{i=1}^{N-1} h_{i} Pr_{i}|Pr_{0}$, where $Pr_{i}|Pr_{0}$ refers to the received power from the $i$-th interfering UAV-BS conditioned on the maximum received power $Pr_{0}$ from the serving UAV-BS. 

\subsection{Performance Analysis}
In this subsection, the coverage probability is analytically investigated based on the following definition $\mathcal{P}_{c}(\theta) \triangleq  \mathbb{P}({\rm{SIR}} > \theta)$, where $\theta$ denotes an SIR threshold value. To this end, the PDF of the maximum received power is first obtained. 

\emph{Lemma 1.} \emph{The PDF of the maximum received power $Pr_0$ from the serving UAV-BS is given by} 
\begin{equation}
f_{Pr_0}(x_0) = N (F_{Pr_{i}}(x_0))^{N-1} f_{Pr_{i}}(x_0), 
\end{equation} 
\emph{where $f_{Pr_{i}}(x)$ denotes the PDF of $Pr_{i}$  given by}
\begin{equation}
f_{Pr_{i}}(x) = \int_{\sqrt{h^2+R^2}^{-\alpha}}^{h^{-\alpha}} \frac{1}{w} f_{l(d_i)}(w) f_{S_i}(x/w) {\rm{d}} w, 
\end{equation}
\emph{ for $x \in [0,\infty)$ and $f_{l(d_i)}(x)$ is the PDF of $l(d_i)$ given by  
\begin{equation}
f_{l(d_i)}(x) = \frac{1}{\alpha R} \frac{x^{-\frac{\alpha+2}{\alpha}}}{\sqrt{x^{-\frac{2}{\alpha}}- h^2}},
\end{equation}
for $x \in [\sqrt{h^2+R^2}^{-\alpha}, h^{-\alpha}]$, and $F_{Pr_{i}}(x_0)$ denotes the CDF of $Pr_{i}$ given by $F_{Pr_{i}}(x_0)= \int_0^{x_0} f_{Pr_{i}}(x) {\rm{d}} x$}.

\textit{Proof.} See Appendix A. \QEDA

Having statistically characterized the PDF of the maximum received power, the conditional Laplace transform of the aggregate interference power distribution can now be derived.

\emph{Lemma 2.} \emph{Conditioned on the maximum received power $Pr_0 = x_0$, the  conditional Laplace transform  $\mathcal{L}_{I}(s|x_0)$ of the aggregate interference power distribution is given by}
\begin{equation}
\mathcal{L}_{I}(s|x_0) = \Big[\int_{0}^{x_0} \Big(1+\frac{s p_i}{m}\Big)^{-m} f_{Pr_{i}|x_0}(p_i) {\rm{d}} p_i\Big]^{N-1}, 
\end{equation}
\emph{where $f_{Pr_{i}|x_0}(p_i)$ denotes the conditional PDF of the received power from the $i$-th interfering UAV-BS conditioned on $x_0$ and is given by}
\begin{equation}
f_{Pr_{i}|x_0}(p_i) = \frac{f_{Pr_{i}}(p_i)}{F_{Pr_{i}}(x_0)}.  
\end{equation}

\textit{Proof.} See Appendix B. \QEDA

\emph{Theorem 1.} \emph{The coverage probability of a receiver in a UAV corridor-assisted network under the maximum power-based association policy is given by}
\begin{align}
&\mathcal{P}_{c}(\theta) =\\
& \int_{0}^{\infty} \sum_{k=0}^{m - 1} \frac{{(-m \theta x_0)}^k }{k!} {\Bigg[ \frac{\partial^{k} \mathcal{L}_{I}(s|x_0)}{\partial{s^k}}\Bigg]}_{s = m \theta  x_0} f_{Pr_0}(x_0) {\rm{d}} x_0.  \nonumber 
\end{align}

\textit{Proof.} The proof for the derivation of Theorem 1 follows steps similar to the proof for deriving Theorem 1 in \cite{TWC}, and hence it is omitted here. \QEDA

\subsection{Special Cases: A Dominant Interferer Approach}

In this section, coverage probability analysis is conducted under a dominant interferer assumption\footnote{Note that the dominant interferer approach avoids the analytical complicated calculation of the derivative of the Laplace transform in Theorem 1, especially for the larger values of $m$, while the accuracy is maintained.}. Accordingly, the effect of the dominant interferer is exactly captured, while the aggregate interference power from the rest of the interfering UAV-BSs is approximated by its mean. As a special case, the coverage
performance is also investigated under the assumption of neglecting all but a single dominant interferer. By considering maximum power user association policy, the receiver associates with the UAV-BS providing the maximum received power as described in subsection III-A-1. Clearly, the dominant interferer is the UAV-BS that provides the second largest received power after the serving UAV-BS. The received SIR is then given by 
\begin{equation}
\begin{split}
{\rm{SIR}}  &= \frac{h_0 Pr_{0}}{h_I Pr_I + I_{N-2}}  \approx \frac{h_0 Pr_{0}}{h_I Pr_I + \Omega(Pr_{0},Pr_I)},
\end{split}
\end{equation}
where $I_{N-2} = \sum_{i=1}^{N-2} h_i Pr_i|(Pr_{0},Pr_I)$,  $\Omega(Pr_{0},Pr_I)$ denotes the conditional mean of the interference power at the receiver conditioned on $Pr_0$ and excluding the interference from the dominant interfering UAV-BS and it is given by $\Omega(Pr_{0},Pr_I) = \mathbb{E}[I_{N-2}]$.  Note that conditioned on $Pr_0$ and excluding interference from the dominant interfering UAV-BS, the $N-2$ RVs of $I_{N-2}$ are conditionally independent and identically distributed (i.i.d.). Subsequently, since the conditions for applying the Law of Large Numbers are satisfied, the approximation in (12) is valid. Moreover, $h_I,Pr_I$ denote the channel power gain and the received power from the dominant interfering UAV-BS, respectively.

To derive the coverage probability, $\Omega(Pr_{0},Pr_I)$ is first obtained. 

\emph{Lemma 3.} \emph{The conditional mean of the interference power at the receiver, excluding the interference from the dominant interfering UAV-BS and conditioned on $Pr_0$, is given by }
\begin{equation}
\Omega(Pr_{0},Pr_I) = (N-2) \int_0^{Pr_I} p_i f_{Pr_{i}|Pr_I}(p_i)  {\rm{d}} p_i,
\end{equation}
\emph{where $f_{Pr_{i}|Pr_I}(p_i)$ denotes the conditional PDF of the received power from the $i$-th interfering UAV-BS conditioned on $Pr_I$ and is given by}
\begin{equation}
f_{Pr_{i}|Pr_I}(p_i) = \frac{f_{Pr_{i}}(p_i)}{F_{Pr_{i}}(Pr_I)}.
\end{equation}

\textit{Proof.}  See Appendix C. \QEDA

Next, the joint PDF of $f_{Pr_{0},Pr_{I}}(x_0,x_I)$ is derived as an intermediate step in the coverage probability. 

\emph{Lemma 4.} \emph{The joint PDF of the maximum received power $Pr_0$ from the serving BS and the second largest power $Pr_I$ from the dominant interferer is given by} 
\begin{equation}
\begin{split}
&f_{Pr_{0},Pr_{I}}(x_0,x_I) \\
&= N (N-1) f_{Pr_{i}}(x_0) f_{Pr_{i}}(x_I) (F_{Pr_{i}}(x_I))^{N-2}.
\end{split}
\end{equation}

\textit{Proof.} The proof follows directly using the results of order statistics \cite{orderstatistics} for the joint PDF of the two largest RVs; hence, a detailed proof is omitted here. \QEDA

The coverage probability can now be obtained as follows. 

\emph{Proposition 1.} \emph{The coverage probability of a receiver in a UAV corridor-assisted network under a dominant interferer approach is given by} 
\begin{equation}
\begin{split}
&\mathcal{P}_{c}(\theta) = \int_{0}^{\infty} \int_{0}^{\infty} \int_{0}^{x_0}  \frac{\Gamma\Big(m,m \frac{\theta h x_I + \theta \Omega(x_0,x_I) }{x_0}\Big)}{\Gamma(m)}\\
& \times f_{Pr_{0},Pr_{I}}(x_0,x_I) f_{h_i}(h) {\rm{d}} x_I {\rm{d}} x_0  {\rm{d}} h,  
\end{split}
\end{equation}
\emph{where $\Gamma(\cdot,\cdot)$ denotes the upper incomplete  gamma function defined as in \cite[eq. (8.350.2)]{Ryzhik}.} 

\textit{Proof.} See Appendix D. \QEDA

By neglecting all but the single dominant interfering UAV-BS, the coverage probability is further simplified as follows. 

\emph{Corollary 1.} \emph{The coverage probability of a receiver in a UAV corridor-assisted network in the presence of a single dominant interfering UAV-BS is given by} 
\begin{align}
&\mathcal{P}_{c}(\theta) = \int_{0}^{\infty} \int_{0}^{\infty} \int_{0}^{x_0}  \frac{\Gamma\big(m,m \frac{\theta h x_I }{x_0}\big)}{\Gamma(m)} f_{Pr_{0},Pr_{I}}(x_0,x_I)  \nonumber \\
&\times f_{h_i}(h) {\rm{d}} x_I {\rm{d}} x_0  {\rm{d}} h.  
\end{align}

\section{Finite HPPP Performance Analysis}
\subsection{User Association Policy and SIR Definition}

\subsubsection{User Association Policy} Similar to subsection \ref{sub:user_association}, by considering a maximum power user association policy, the receiver associates with the UAV-BS providing the maximum average received power within $\Phi$. Accordingly, the maximum received power is given by
\begin{equation}
Pr_{0} = \underset{\begin{subarray}{c}
  y \in \Phi
  \end{subarray}}{\operatorname{max}}\{Pr_y\} \\ 
  = \underset{y \in \Phi} {\operatorname{max}}\{{S_y}\, l(d_y)\}.   
\end{equation}
The location of the serving UAV-BS is chosen by the receiver as
\begin{equation}
y_0 =  \underset{y \in \Phi} {\operatorname{argmax}}\{{S_y}\, l(d_y)\}.   
\end{equation}
Similarly to BPP-based spatial modeling of UAV-BSs, the serving UAV-BS in this scenario is not necessarily the closest in terms of Euclidean distance.

\subsubsection{SIR Definition}
Assuming that all UAV-BSs share the same resource blocks, then $N-1$ UAV-BSs interfere the receiver. The received SIR is given by 
\begin{equation}
{\rm{SIR}} = \frac{h_{0} Pr_{0}}{I}, 
\end{equation}
where $I$ refers to the aggregate interference power and is given by $I = \sum_{y \in \Phi^{!}} h_{y} S_{y} l(d_y)$, where $ \Phi^{!} =  \Phi \setminus \{y_0\}$. 

\subsection{Performance Analysis}
In this section, a coverage probability analysis is performed. To this end, the PDF of the maximum received power from the serving UAV-BS is first obtained. 

\emph{Lemma 5.} \emph{Conditioned on the event $\mathcal{A}_{L(-R,R)}$, the PDF of the maximum received power $Pr_0$ from the serving UAV-BS is given by } 
\begin{align}
&f_{Pr_0|\mathcal{A}_{L(-R,R)}}(x_0)\\
&= \frac{\lambda_{UAV} |L(-R,R)|f_{Pr_{y}}(x_0)e^{\lambda_{UAV} |L(-R,R)| (F_{Pr_{y}(x_0)}-1)}}{\mathbb{P}[\mathcal{A}_{L(-R,R),1}]},   \nonumber
\end{align}
\emph{where $f_{Pr_{y}}(x)$ denotes the PDF of $Pr_{y}$  given by}
\begin{equation}
f_{Pr_{y}}(x) = \int_{\sqrt{h^2+R^2}^{-\alpha}}^{h^{-\alpha}} \frac{1}{w} f_{l(d_y)}(w) f_{S_y}(x/w) {\rm{d}} w, 
\end{equation}
\emph{and $\mathcal{A}_{L(-R,R)}$ denotes the event that $L(-R,R)$ has at least $1$ UAV-BS of $\Phi$ and therefore $\mathbb{P}[\mathcal{A}_{L(-R,R)}]$ is given by $\mathbb{P}[\mathcal{A}_{L(-R,R)}]= 1- e^{-\lambda_{UAV} |L(-R,R)|}$.} 

\textit{Proof.} See Appendix E. \QEDA

\begin{figure}[!t]
    \centering
    \includegraphics[keepaspectratio,width= \linewidth]{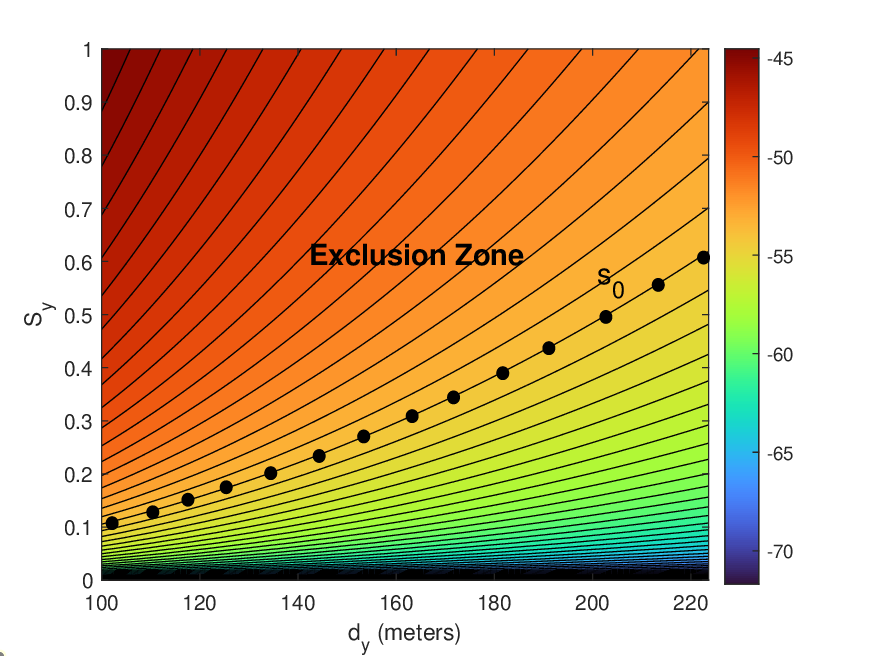}
       \caption{The geometric locus of the points with received power $Pr_y =  S_y d_y^{-\alpha}$, versus $S_y$ and $d_y$ given a maximum received power $s_0$}
\end{figure}

\emph{Lemma 6.} \emph{Conditioned on the maximum received power $Pr_0=s_0$ from the serving UAV-BS, the conditional Laplace transform $\mathcal{L}_I(s|s_0)$  of
the aggregate interference power distribution is given by \eqref{eqLT}, shown at the bottom of the next page, where} $\mathbf{\Omega} = \{ y \in \Phi^! | h  \leq d_y \leq  \sqrt{h^2+R^2}, \, 0 \leq s_y \leq s_0 d_y^{\alpha}\}$. 

\begin{figure*}[!b]
\hrulefill
\begin{equation} \label{eqLT}
\begin{split}
\mathcal{L}_I(s|s_0) = {\rm{exp}}\Bigg(- 2 \lambda_{UAV} \iint_{\mathbf{\Omega}} \Big(1- \Big(1+\frac{s\, s_y\,  d_y^{-\alpha_L}}{m} \Big)^{-m}\Big)\frac{d_y}{\sqrt{d_y^2+R^2}}f_{S_y}(s_y)  {\rm d} s_y {\rm d} d_y  \Bigg),
\end{split}
\end{equation}
\end{figure*}

\begin{figure}[!t]
    \centering
    \includegraphics[keepaspectratio,width= 0.9\linewidth]{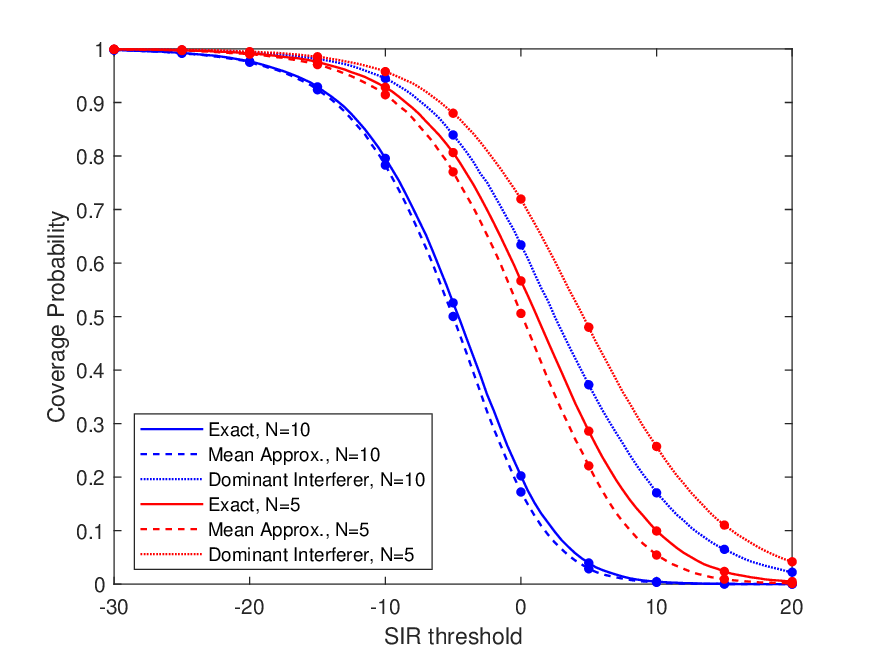}
       \caption{Coverage probability, versus threshold $\theta$ under exact and dominant interferer approaches, for two values of $N$ UAV-BSs. Markers denote analytical results.}
\end{figure}

\begin{figure}[!t]
    \centering
    \includegraphics[keepaspectratio,width= 0.9\linewidth]{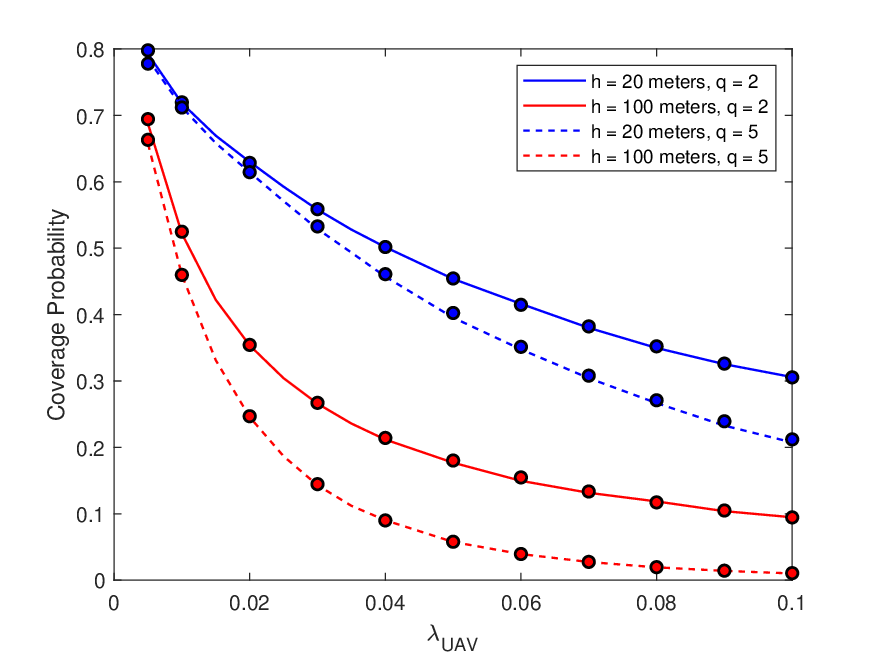}
       \caption{Coverage probability versus $\lambda_{UAV}$ for different values of $h$ and $q$. Markers denote analytical results.}
\end{figure}

\begin{figure}[!t]
    \centering
    \includegraphics[keepaspectratio,width= 0.9\linewidth]{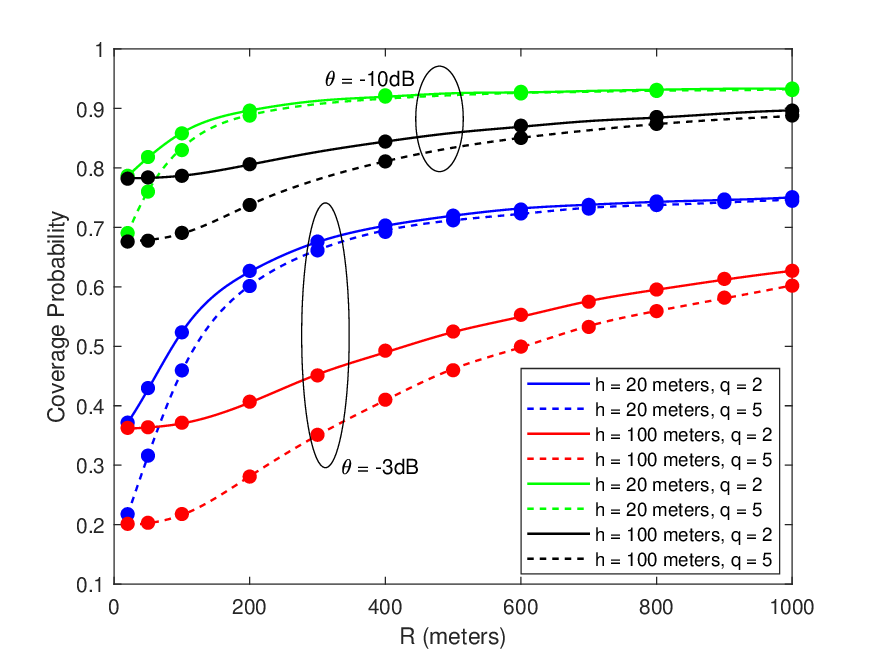}
       \caption{Coverage probability versus $R$ for two values of $\theta$ and different values of $h$ and $q$. Markers denote analytical results.}
\end{figure}

\begin{figure}[!t]
    \centering
    \includegraphics[keepaspectratio,width= 0.9\linewidth]{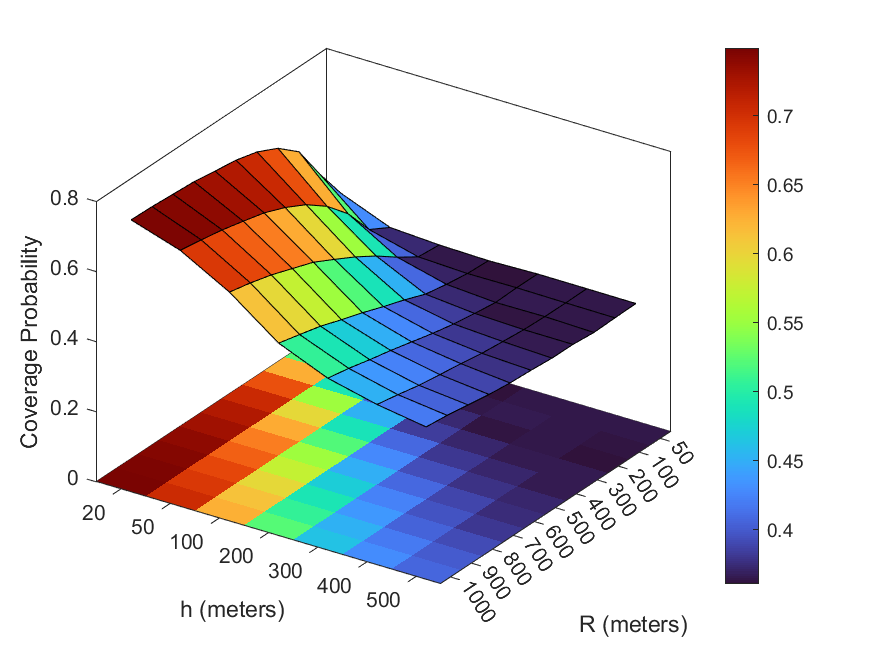}
       \caption{Coverage probability versus $h$ and $R$ for $\theta = -3$ dB.}
\end{figure}

\textit{Proof.} Conditioned on $Pr_0 = s_0$, the location $y_0$ and the shadowing conditions of the serving UAV-BS is known (refer to (19)). Once the serving UAV-BS has been determined, appealing to Slivnyak's theorem, $\Phi^{!}$ is also a finite HPPP with density $\lambda_{UAV}$. Next, an exclusion zone is defined by all possible distances and shadowing conditions w.r.t. $y_0$ leading to the maximum received power $s_0$. Accordingly, the set $\mathbf{\Omega}$ is defined by all possible locations of the interfering UAV-BSs for which the distances $d_y$ and shadowing conditions $S_y$ lie outside the exclusion zone. However, the location $y_0$ of the serving UAV-BS is a function of its maximum received power $s_0$. In Fig. 2, a representative example of the maximum received power $s_0 = -54$dBm from the UAV-BS at $y_0$ is presented. More specifically, the geometric locus of the points with received power $Pr_y= S_y d_y^{-\alpha}$ from $ y \in \Phi^!$ in terms of $(d_y,S_y)$, is shown. Note that for a given value of $Pr_0=s_0$ in Fig. 2, the received power from the interfering UAV-BSs should be smaller than $s_0$, and all $y \in \Phi^{!}$ must lie outside the exclusion zone indicated by the black markers. Another interesting observation in Fig. 2 is that $s_0$ also determines the maximum received power of an interfering UAV-BS. Now conditioned on $s_0$, the large-scale fading  $S_y$ of the interfering UAV-BSs is a function of their distances $d_y$ and is given by
\begin{equation} 
S_y d_y^{-\alpha} <s_0 \Leftrightarrow S_y < s_0 d_y^{\alpha}, 
\end{equation}
as $d_y \in [h,\sqrt{h^2+R^2}]$. Next, the Laplace transform can be obtained as described in Appendix F. \QEDA 


\emph{Theorem 2.} \emph{Conditioned on $\mathcal{A}_{L(-R,R)}$ the coverage probability of a receiver in a UAV corridor-assisted network under
the maximum power-based association policy is given by}
\begin{equation}
\begin{split}
&\mathcal{P}_{c}(\theta) = \int_{0}^{\infty} \sum_{k=0}^{m - 1} \frac{{(-m \theta s_0)}^k }{k!} {\Bigg[ \frac{\partial^{k} \mathcal{L}_{I}(s|s_0)}{\partial{s^k}}\Bigg]}_{s = m \theta  s_0} \\
&\times f_{Pr_0|\mathcal{A}_{L(-R,R)}}(s_0) {\rm{d}} s_0.  
\end{split}
\end{equation}

\textit{Proof.} The proof for the derivation of Theorem 2 follows the same conceptual lines as the proof for the derivation of Theorem 1 with the only difference being the deconditioning with the PDF $f_{Pr_0|\mathcal{A}_{L(-R,R)}}(s_0)$. \QEDA

\section{Numerical Results, Experimental Validation \& Discussions}
In this section, numerical results are presented to evaluate and compare the performance achieved in a UAV corridor-assisted network under the finite HPPP and BPP spatial modeling of UAVs. The accuracy of the analytical results is verified by comparing them with the simulated results obtained using Monte Carlo simulations. Moreover, experimental results are also included and compared with simulated ones to validate the proposed concept of the maximum power association policy and, subsequently, the coverage probability analysis.

\subsection{Numerical Results}
For all numerical results, the following parameters have been assumed unless otherwise stated: $N=10$, $\lambda_{UAV}=N/|L(-R,R)|$, $m=1$  in order to capture the worst-case fading scenario, $q=2$ \cite{Bithas}, $a=2.2$ \cite{9893878} which was also shown to match with our experimental results, $h=100$ meters \cite{9785498}, $R=500$ meters \cite{7936620} and $\theta = -3$dB.

Fig. 3 presents the coverage probability for the BPP spatial modeling of UAV-BSs, obtained by Theorem 1, versus the SIR threshold $\theta$  for two values of UAV-BSs $N$. The coverage probabilities under dominant interferer approaches, given by Proposition 1 and Corollary 1, are also shown. It is observed that by reducing the number of UAV-BSs deployed, the probability of coverage increases as the aggregate interference power is reduced. Moreover, the approximation of the mean interference power is accurate for all values of $N$, while the single dominant interferer approach becomes more accurate only for small values of $N$. 

Fig. 4 presents the coverage probability for the case of finite HPPP spatial modeling of UAV-BSs, obtained by Theorem 2, versus the deployment density of UAV-BSs $\lambda_{UAV}$ for different values of $h$ and shadowing conditions $q$. The first observation is that, under the same shadowing conditions (e.g., $q=2$), the coverage probability decreases as $h$ increases. In contrast to this, as the shadowing conditions are getting worse (from $q=2$ to $q=5$), the decrease in coverage performance is more profound for higher UAV deployment heights $h$. 

Fig. 5 presents the coverage probability for the case of finite HPPP spatial modeling of UAV-BSs, obtained by Theorem 2, versus $R$ for two values of the SIR threshold $\theta$ and different values of $h$ and $q$. It is observed that for a given value of $h$, the gap between coverage performance under light and severe shadowing conditions tends to diminish for UAV corridors with large dimensions. However, it is interesting to observe that the convergence is significantly faster for smaller heights of deployment of UAVs $h$. 

Fig. 6 presents the simulated coverage probability versus $R$ and $h$ for the SIR threshold $\theta=-3$ dB. Interestingly, notice that given a value of the length of the UAV corridor $2 R$, an increase in the height of deployment of the UAV $h$ causes the probability of coverage to rapidly decay. In contrast, given a value of the height of deployment of the UAV $h$, a decrease in $R$  makes the coverage performance decrease more smoothly. Finally, notice that the maximum coverage probability is achieved for large UAV corridors when deployed at low heights.  Indeed, by employing large UAV corridors at low heights, the aggregate interference is reduced due to two main reasons: i) The low deployment height worsens the interfering UAV-BSs' channel conditions and ii) the distance between the serving UAV-BS and the interfering UAV-BSs tends to increase.

\begin{figure}[!t]
    \centering
    \includegraphics[keepaspectratio,width= 0.9\linewidth]{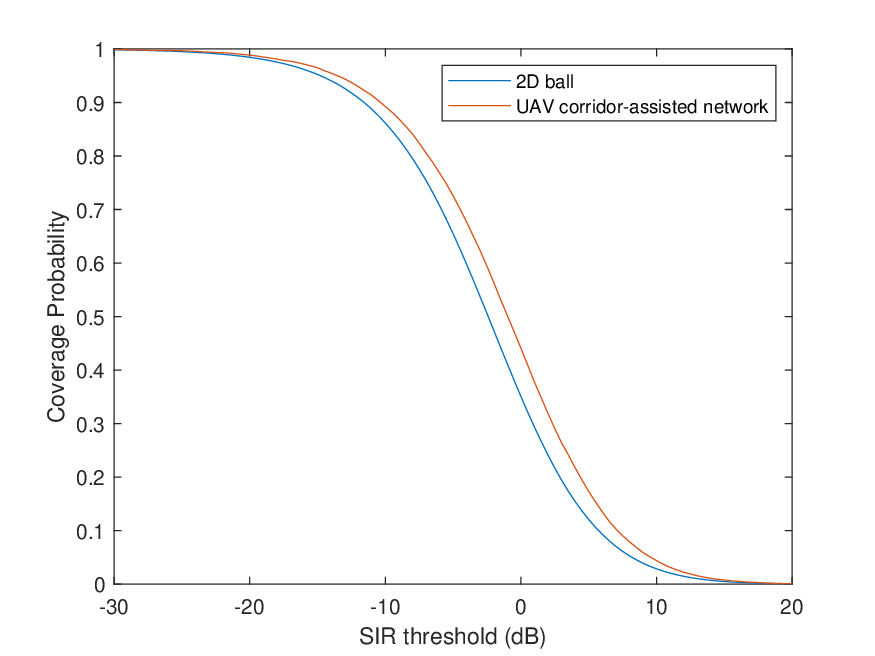}
       \caption{Performance comparison between the coverage probability under the UAV corridor-assisted network and a UAV network whose UAV-BSs are deployed in a 2D disc above the ground under the BPP spatial modeling of UAV-BSs.}
        \label{BPP}
\end{figure}

Finally, in Fig. \ref{BPP} a performance comparison is performed between the UAV corridor-assisted network and a UAV network whose UAV-BSs are deployed in a 2D ball $b(\mathbf{o},R)$ centered at the origin $\mathbf{o}$ with radius $R$ for the representative case of the BPP spatial modeling of the UAV-BSs. Note that the 2D ball-based spatial model is extremely common in the open technical literature \cite{Dhillon,10050345}. The results are shown in Fig. \ref{BPP}. For the key simulation parameters, we assume $h=50$ meters, $R = 250$ meters, and $N=10$ UAV-BSs. Interestingly, in Fig. \ref{BPP} we observe that the coverage performance in the UAV corridor-assisted UAV network clearly outperforms the corresponding one in a UAV network whose UAV-BSs are deployed as a BPP in a 2D disc above the ground. This is because of the spatial model's geometry, i.e., the probability that more interfering UAV-BSs hover above the receiver is higher under the 2D ball model as compared to the one under the UAV corridor-assisted network.

\begin{figure*}[!htbp]
  \begin{subfigure}[t]{.32\linewidth}
  \includegraphics[trim=20 5 20 15,clip,width=\linewidth]{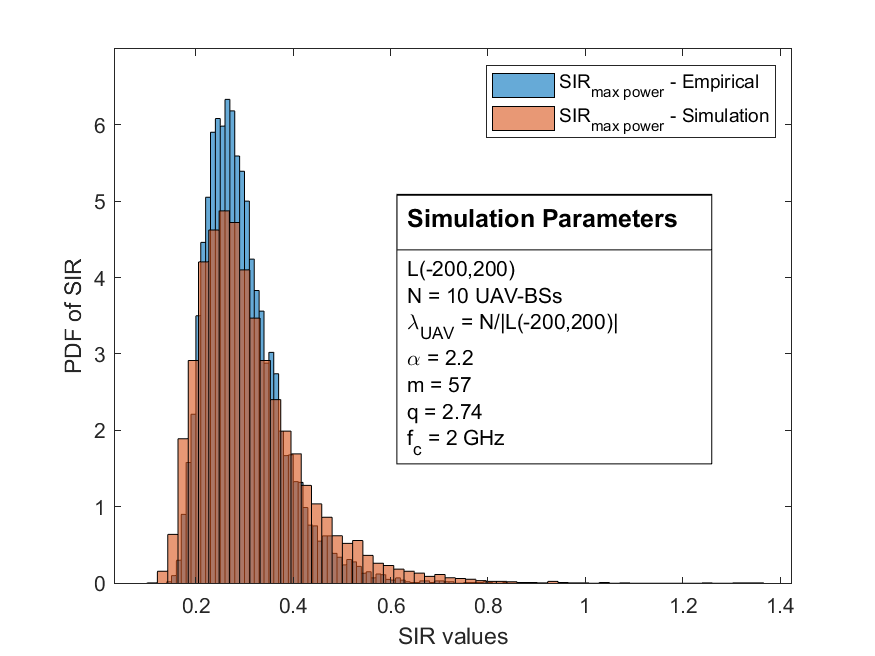}
    \caption{Empirical versus simulated PDF of SIR under the max. power association policy.}%
\label{EmpiricalBPPa}
  \end{subfigure}\hfil
  \begin{subfigure}[t]{.32\linewidth}
    \includegraphics[trim=20 5 20 15,clip,width=\linewidth]{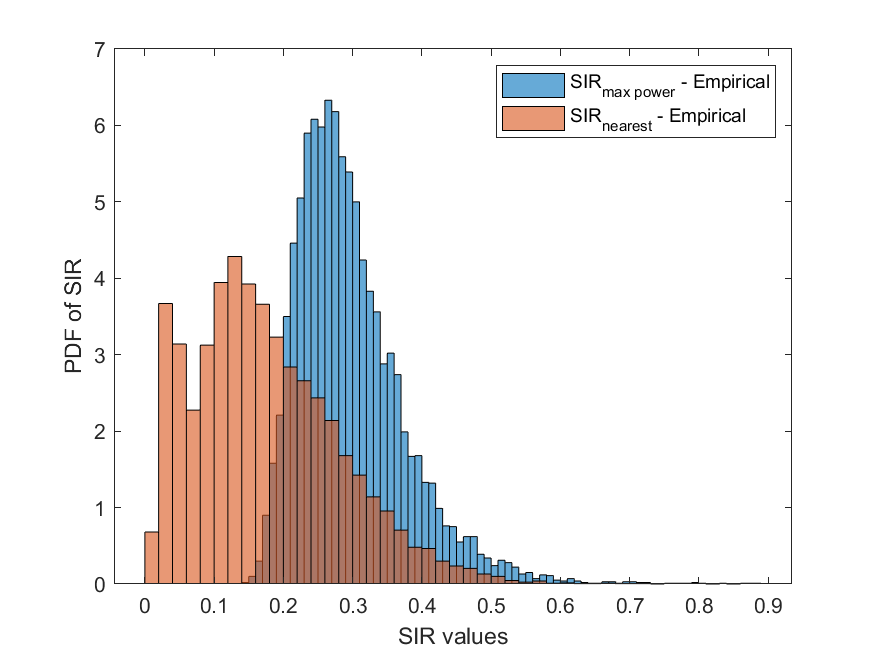}%
    \caption{Empirical PDF of SIR under the max. power versus the min. distance association policy.}%
     \label{EmpiricalBPPb}
  \end{subfigure}\hfil
  \begin{subfigure}[t]{.32\linewidth}
    \includegraphics[trim=20 5 20 15,clip,width=\linewidth]{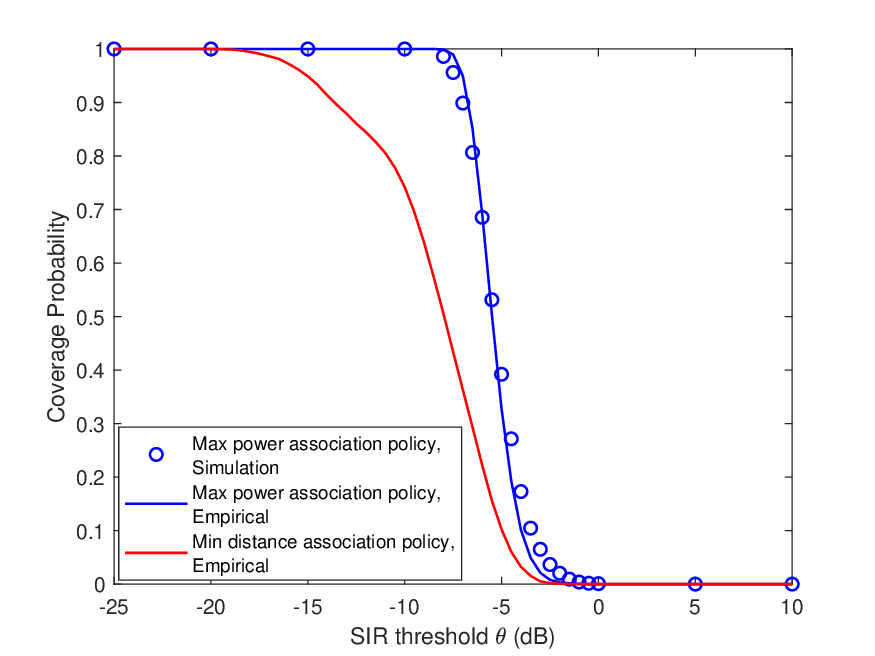}%
    \caption{Empirical versus simulated coverage probability under max. power and min. distance association policy.}%
    \label{EmpiricalBPPc}
  \end{subfigure}%
   \caption{Empirical versus simulated results for BPP spatial modeling of UAVs in the corridor. }
   \label{EmpiricalBPP}
\end{figure*}

\subsection{Experimental Validation \& Comparisons}
In this section, the experimental results obtained in a measurement campaign \cite{Vik} will be presented and compared to the simulated results to verify the accuracy of the theoretical approach in real-world scenarios. The measurement campaign took place in Prague, in an urban pedestrian environment, with the receiver positioned on the street level in the middle of a crossroad, while the transmitter was mounted on the bottom part of a Zeppelin-type airship. The airship moved at a constant speed of $6.2$ m/sec, in a predefined route over the position where the receiver was located. The height above the ground was maintained at approximately $200$ meters while the receiver was kept stationary. The airship was equipped with a GPS sensor, enabling the transmitter to constantly point the antennas towards the receiver by using a positioner attached to the bottom part of the ship and allowing the distance between the receiver and the transmitter to be calculated.  From the data collected, a specific segment in which the airship was flying in a straight line over the ground was selected. In this scenario, the received signal power was analyzed using a significant number of airship position realizations generated according to a BPP or a finite HPPP. These realizations were used to emulate the power received from multiple UAVs. This was feasible since the sampling rate of the measurement system was 10 kHz which, in conjunction with the airship moving speed, resulted in 258 samples/$\lambda$. Therefore, any simulated/generated position can be mapped to a recorded position with an accuracy of $0.5 \, mm$ ($\lambda = 15 \,cm$). This setup allowed us to compare the performance of the proposed BPP or finite HPPP-based spatial UAV distribution with that derived from empirical measurements. The high-level methodology for simulating-an equivalent to the measurement-UAV corridor-assisted scenario, is presented as follows: First, we simulate the spatial locations (both as BPP and as a finite HPPP) of the UAV-BSs across the corridor, and subsequently we repeat the experiment 10000 times (also referred to as trials). For each trial, we find the UAV-BS, referred to as serving UAV-BS, which provides the maximum received power while the received power from all other UAV-BSs is considered as aggregate interference at the receiver. It should be mentioned that information about the power received from the locations of the interfering UAV-BSs across the corridor is available since the airship had run through the whole line segment. More details for the measurement campaign can be found in \cite{Vik}. 

For the calculation of the empirical SIR, the derivation process in a high level is described as follows: From a set of simulated data of $\Phi$ and $\Psi$ with $\lambda_{UAV}=10/|L(-200,200)|$ and $N=10$, respectively, the corresponding locations are extracted from the empirical data. Then, for each of the UAVs' locations, the corresponding received powers at the UE are returned and sorted in descending order. Accordingly, the UE is assumed to be associated with the UAV-BS providing the maximum power, whereas the remaining UAVs for both spatial models are assumed to interfere to the UE. 

Fig. \ref{EmpiricalBPP} presents the empirical versus simulated results for the case of BPP spatial modeling of UAVs. Fig. \ref{EmpiricalBPPa} compares the empirical PDF of SIR under the maximum power association policy versus the simulated one, obtained by following the procedure described in subsection \ref{sub:user_association}. The simulated PDF has been generated using the parameters as shown in the table included in that figure. Notably, the simulated PDF of SIR matches particularly well with the empirical one. This clearly verifies that: i) the formulation of the maximum power association policy, as given by (4), is accurate, and ii) the SIR metric, as given by (5), holds true for real-world UAV application scenarios. Fig. \ref{EmpiricalBPPb} compares the empirical PDFs of the SIR under the maximum power versus the minimum distance receiver association policy. Two key insights can be drawn: i) Attaching to the nearest UAV-BS during the receiver's association policy tends to severely underestimate the SIR metric and ii) the effect of shadowing in UAV-assisted networks has a key role in the SIR metric. Indeed, by considering the oversimplified scenario of neglecting the shadowing effect in the calculation of received power, the maximum power association policy is deduced to the conventional minimum Euclidean distance criterion. Finally, Fig. \ref{EmpiricalBPPc} shows the empirical versus simulated coverage probability under the maximum power and minimum distance association policy. The first observation is that the empirical coverage performance fits well with the simulated one, which verifies the need for a more nuanced modeling of the maximum power association policy. Moreover, the coverage performance under the minimum Euclidean distance criterion is clearly underestimated. 

\begin{figure*}
    \centering
    \begin{subfigure}[b]{0.45\linewidth} 
        \centering
        \includegraphics[trim=15 5 30 15,clip,width=\linewidth]{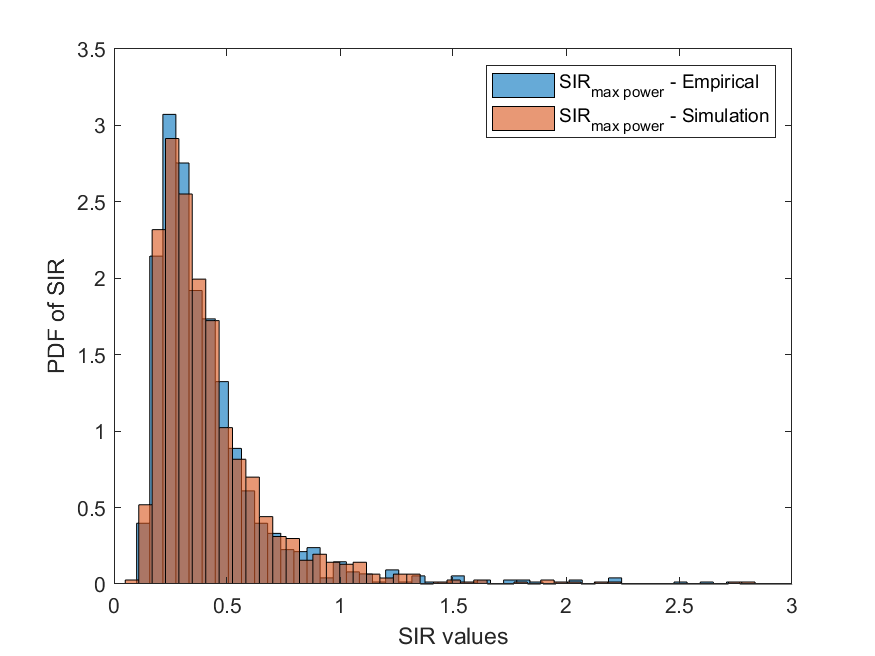}
        \caption{{\small Empirical versus simulated PDF of SIR under the max. power association policy.}}    
       \label{EmpiricalFHPPPa}
    \end{subfigure}
    \hspace{0.01\linewidth} 
    \begin{subfigure}[b]{0.45\linewidth}  
        \centering 
        \includegraphics[trim=15 5 30 15,clip,width=\linewidth]{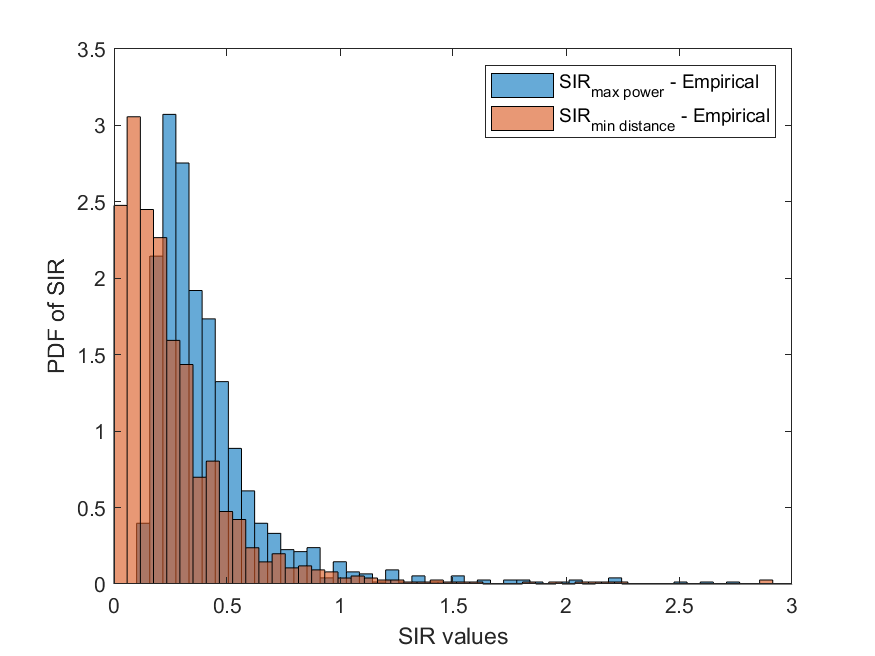}
        \caption{{\small Empirical PDF of SIR under the max. power versus the min. distance association policy.}}    
      \label{EmpiricalFHPPPb}
    \end{subfigure}
    
    \vspace{0.5em} 

    \begin{subfigure}[b]{0.45\linewidth}   
        \centering 
        \includegraphics[trim=15 5 30 15,clip,width=\linewidth]{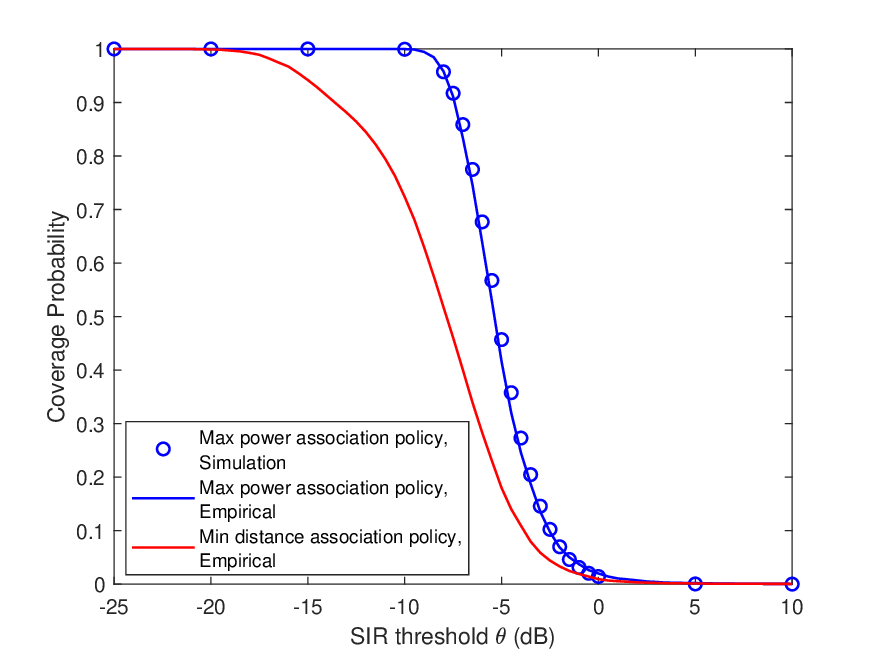}
        \caption{{\small Empirical versus simulated coverage probability under max. power and min. distance association policy.}}    
     \label{EmpiricalFHPPPc}
    \end{subfigure}
    \hspace{0.01\linewidth} 
    \begin{subfigure}[b]{0.45\linewidth}   
        \centering 
        \includegraphics[trim=15 5 30 15,clip,width=\linewidth]{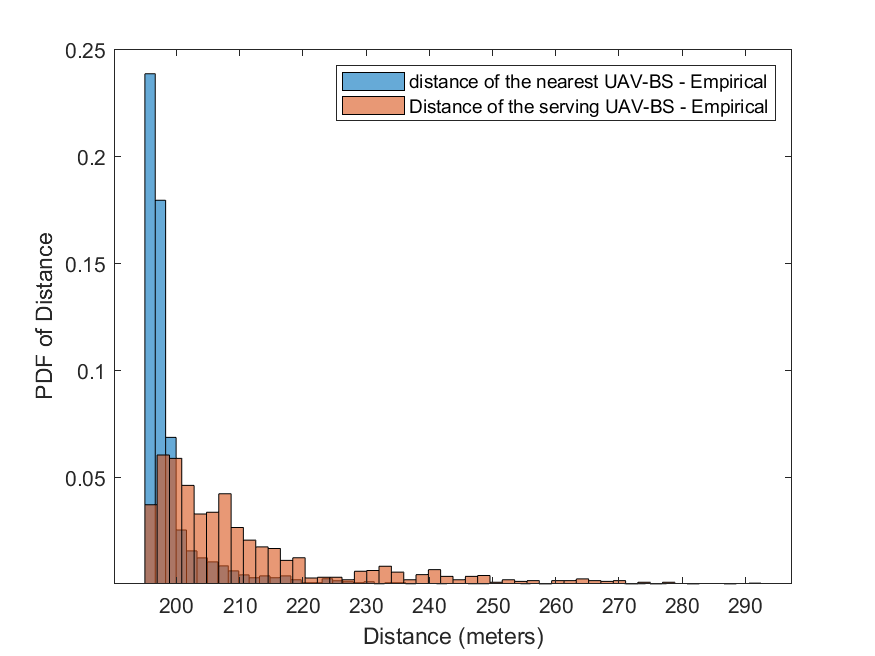}
        \caption{{\small Empirical PDF of the Euclidean distance of the nearest UAV-BS versus the distance of the serving-BS.}}    
        \label{EmpiricalFHPPPd}
    \end{subfigure}
    \caption{Empirical versus simulated results for finite HPPP spatial modeling of UAVs in the corridor.} 
   \label{EmpiricalFHPPP}
\end{figure*}

Fig. \ref{EmpiricalFHPPP} presents empirical versus simulated results for the case of finite HPPP spatial modeling of UAVs. Fig. \ref{EmpiricalFHPPPa} compares the empirical PDF of SIR under the maximum power association policy versus the simulated one. The simulated PDF has been generated using the same parameters shown in the table included in Fig. \ref{EmpiricalBPPa}. It is observed that the simulated PDF of SIR matches almost perfectly with the experimental one, which clearly validates the necessity of capturing the shadowing in the maximum power user association policy. Fig. \ref{EmpiricalFHPPPb} confirms that although the mismatch between the empirical PDFs of SIR under the maximum power and minimum distance association policy is not that large compared to the one in Fig. \ref{EmpiricalBPPb}, it still exists. Fig. \ref{EmpiricalFHPPPc} also verifies the good fitness between the empirical and simulated coverage probability for the finite HPPP spatial model of the UAV-BSs. Finally, Fig. \ref{EmpiricalFHPPPd} shows the empirical PDFs of the distance between the receiver and the nearest UAV-BS and the distance between the receiver and the serving BS providing the maximum power. Quite interestingly, it is clearly observed that the two PDFs differ significantly and present different statistical behavior. This fact also verifies that in real-world UAV scenarios in urban environments, the assumption of associating with the nearest UAV-BS is not accurate.

\subsection{Analysis of the Variable Height Model \& Experimental Validation}
We have analyzed the performance of a UAV corridor-assisted network assuming that all the UAV-BSs are deployed in a 1D line segment at the same fixed height $h$. However, it is not unusual for UAVs to be considered at varying heights in practice. Therefore, it is meaningful to comment on the coverage probability analysis under the assumption of variable heights. In fact, the concept of \emph{variable height model} has already been investigated only by simulations in the open technical literature \cite{Dhillon,8443416,9754556}. Indeed, conducting a performance analysis under the variable height model, although mathematically interesting, yields an extremely difficult, if not impossible, result. This is clearly stated in \cite{Dhillon} and in \cite{8443416}. Indeed, for the BPP spatial modeling of the UAV-BSs for example, the distance $\{d_i\}_{i=1:N} = \Big\{\sqrt{r_i^2+h^2}\Big\}$  between the receiver and the UAV-BSs  will turn $d_i$ into a bivariate RV. Even worse, the PDF of $l(d_i)$ in (8) and subsequently the PDF of the maximum received power in Lemma 1 will now depend on the random height, leading to a two-fold integral just to calculate the expression of the PDF of the maximum received power. Nevertheless, as interestingly shown in both of these works, the trend in coverage performance between the fixed- and the variable-height models exhibits almost identical behavior if one assumes a uniform distribution for the UAVs' height in the random-height model. 

\begin{figure}[!t]
    \centering
    \includegraphics[keepaspectratio,width= 0.9\linewidth]{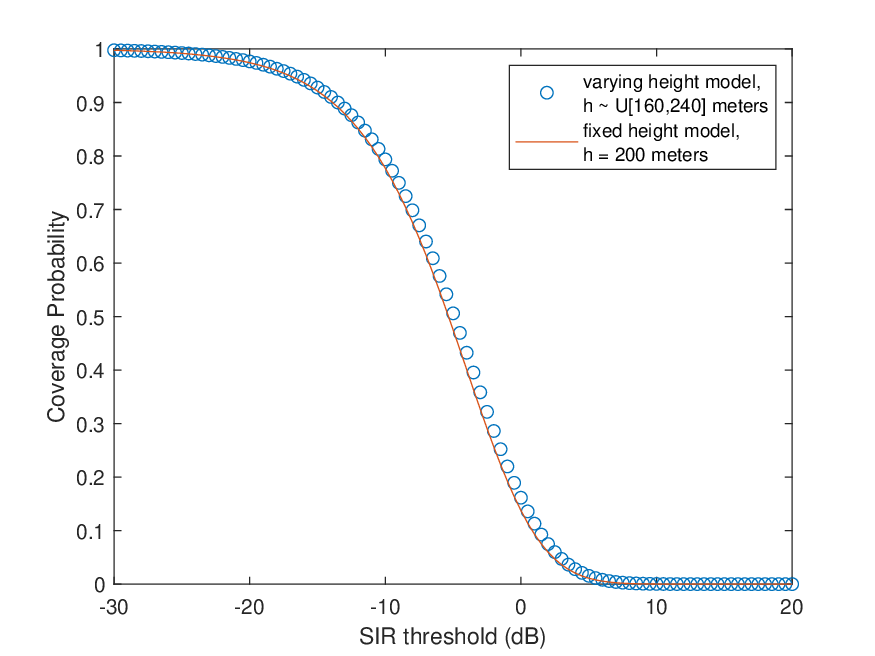}
       \caption{Coverage probability versus SIR threshold under fixed and variable height model for the case of BPP spatial modeling of UAV-BSs.}
        \label{Fig_fit}
\end{figure}

\begin{figure}[!t]
    \centering
    \includegraphics[keepaspectratio,width= 0.9\linewidth]{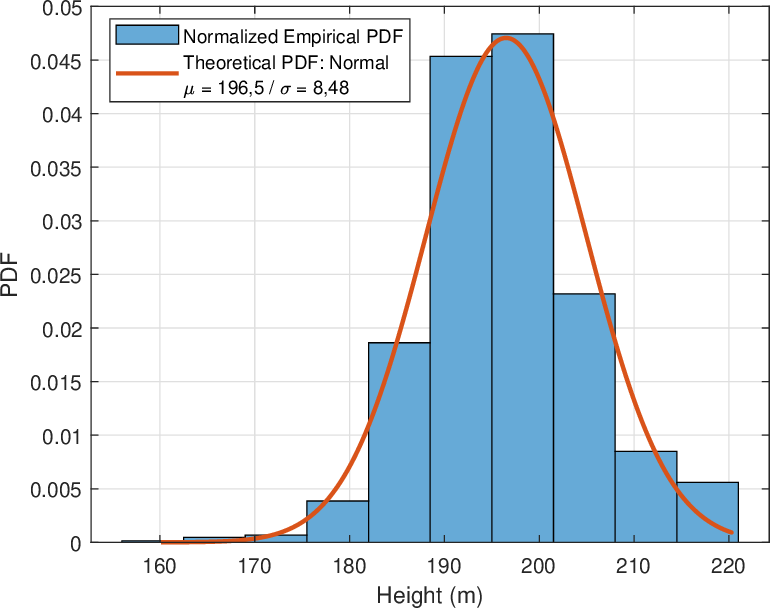}
       \caption{Illustration of empirical PDF of the variable height of the airship versus the Normal distribution fitting.}
        \label{Fig_PDFNormal}
\end{figure}

\begin{figure}[!t]
    \centering
    \includegraphics[keepaspectratio,width= 0.9\linewidth]{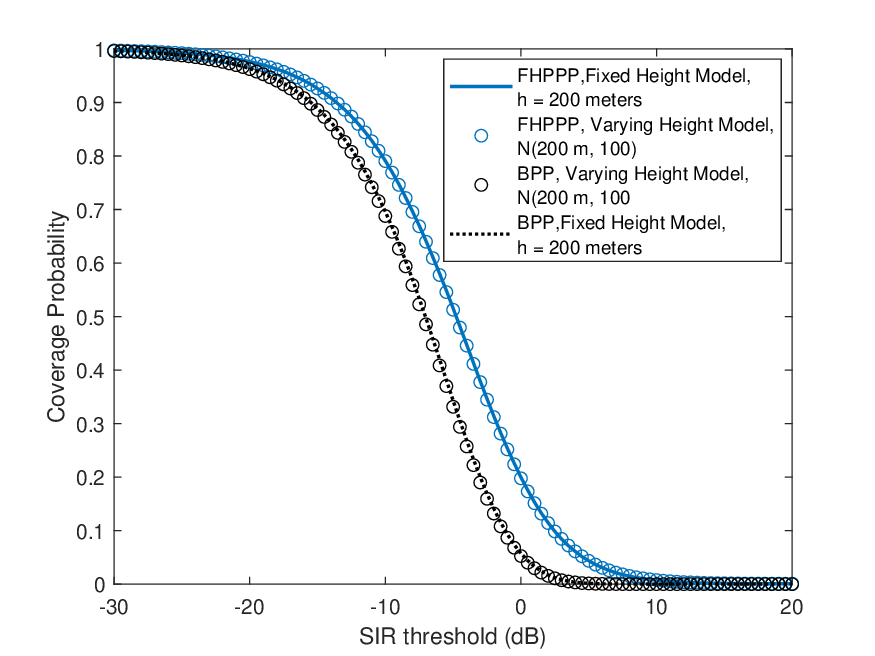}
       \caption{Performance comparison between the coverage probability under fixed and variable height model under Normal distribution for the case of BPP and finite HPPP spatial modeling  of UAV-BSs.}
        \label{Fig_fitNormal}
\end{figure}

Triggered by the aforementioned, in Fig. \ref{Fig_fit}, we validate the coverage performance of our proposed UAV corridor-assisted network under the fixed and varying height model for the representative case of BPP spatial modeling of the UAV-BSs. For the case of fixed height $h$ we assume $h = 200$ m, while for the random height model we assume that the heights of the UAV-BSs are uniformly distributed in the range [160 m, 240 m]. It is clearly observed that the coverage performance under the fixed height model matches well with the one under the variable height model. Until now, we have validated that the observations for the varying heights of the UAVs drawn in \cite{Dhillon,8443416} hold for our UAV corridor-assisted network in which the UAV corridor is modeled as a 1D line segment. However, since the information for the varying height of the airship is available from our empirical data, the authors conduct a statistical analysis to find a reasonable distribution for modeling the varying height. In Fig. \ref{Fig_PDFNormal}, we illustrate the PDF of the airship's variable height extracted from the empirical data. Accordingly, it is shown that the empirical PDF  can accurately be fitted with the Normal distribution. Indeed, in real-world applications, it is not unusual for the height of the UAVs to vary around a mean deployment height. Therefore, \emph{we adopt the Normal distribution as the most reasonable candidate distribution for capturing the random variations of the variable height model.} Note that, to the best of the authors' knowledge, this is the first time that a proposed height distribution for modeling the varying height of UAVs has been verified experimentally. Subsequently, in Fig. \ref{Fig_fitNormal}, we illustrate the coverage probability under fixed and variable height models with normal distribution for the case of BPP and finite HPPP spatial modeling of UAV-BSs. Notably, it is clearly observed both that for both spatial models the coverage performance under the fixed height model matches perfectly with the one under the variable height model. In fact, \emph{the Normal distribution is a more reasonable distribution for modeling the UAVs' varying height as compared to the uniform distribution.} Indeed, using the Kullback-Leibler (KL) divergence criterion to quantify the difference in performance under uniform and normal distributions, we notice that by using the uniform distribution we get a value of 0.000134 and 0.000007 for the BPP and the finite HPPP spatial model, respectively. In contrast, by employing the Normal distribution we get a value of 0.0000086 and 0.0000042 for the BPP and the finite HPPP spatial model, respectively. 

\section{Conclusion}
In this work, a novel stochastic geometry framework based on 1D BPP and 1D finite HPPP was proposed to model the spatial locations of UAV-BSs for UAV corridor-assisted networks. Moreover, since shadowing has an important impact on the coverage performance, large-scale fading has been included in the analysis of user association. Subsequently, coverage probability analysis was conducted for the two spatial models and exact-form expressions were derived. The main results of the work were validated through empirical data collected from the measurement campaign. As a key insight, it was shown that if the shadowing is ignored at the user's association scheme, the coverage performance is significantly underestimated. Among several system-level insights, it was shown that coverage performance is maximized for large UAV corridors deployed at low heights above the ground. As key system-level insights, the results revealed that i) the UAV's altitude distribution was experimentally verified that it can be modeled by a Normal distribution, effectively capturing the random variations in the UAV's flight height and ii) the shadowing conditions of the UAV-BSs, which may significantly vary depending on the city's deployment area, strongly affect the receiver's association policy and subsequently its performance.

\appendices
\section{Proof of Lemma 1}
The PDF $f_{Pr_i}(x)$ of the received power $Pr_i = S_i l(d_i)$ is derived first. Notice that each element $\{Pr_i\}_{i=1:N}$ is an unordered i.i.d RV and can be expressed as a product of two independent RVs. Accordingly, the PDF of $Pr_i$ is given by 
\begin{equation}\label{AppA1}
f_{Pr_{i}}(x) = \int_{\sqrt{h^2+R^2}^{-\alpha}}^{h^{-\alpha}} \frac{1}{w} f_{l(d_i)}(w) f_{S_i}(x/w) {\rm{d}} w,  
\end{equation}
where integration is conducted w.r.t. the range of the RV $l(d_i)$. In \eqref{AppA1}, the PDF of $S_i$ is given directly by (2), while the PDF of $l(d_i) = d_i^{-\alpha}$ is derived as follows. First, the PDF of $d_i$ is given by \cite{frontiers} $f_{d_i}(d) = \frac{d}{R \sqrt{d^2-h^2}}$, for $d \in [h,\sqrt{h^2+R^2}]$. Next, applying the change of RV $d_i = w^{-1/\alpha}$, PDF $f_{l(d_i)}(w)$ can be obtained as in (8). Having obtained the PDF $f_{Pr_{i}}(x)$, the CDF of $F_{Pr_0}(x_0)$ is obtained by exploiting the results of the order statistics \cite{orderstatistics}. Accordingly,  
\begin{align}\label{AppA2}
F_{Pr_0}(x_0) &= \mathbb{P}[Pr_0 \leq x_0] \\
&= \mathbb{P}[Pr_1\leq x_0 ,...,Pr_N\leq x_0] = (F_{Pr_i}(x_0))^N, \nonumber
\end{align}
where $F_{Pr_i}(x_0)$ denotes the CDF of $Pr_i$. Finally, the PDF $f_{Pr_0}(x_0)$ can be obtained as $f_{Pr_0}(x_0) = \frac{d F_{Pr_0}(x_0)}{d x_0} = N (F_{Pr_{i}}(x_0))^{N-1} f_{Pr_{i}}(x_0)$, which results in Lemma 1. 

\section{Proof of Lemma 2}
By applying results from order statistics \cite{orderstatistics} and following conceptual lines similar to those of \cite[Lemma 2]{9740446}, $f_{Pr_{i}|x_0}(p_i)$ is given through the joint PDF of the ordered received powers as $f_{Pr_{i}|x_0}(p_i) = \frac{f_{Pr_{i}}(p_i)}{F_{Pr_{i}}(x_0)}$. Next, conditioned on $x_0$, the conditional Laplace transform of the interference power received from the interfering UAV-BS is given by 
\begin{align} \label{AppB1}
&\mathcal{L}_{I}(s|x_0) = \mathbb{E}_{I}[\exp(-s I)] = \nonumber \\
&= \mathbb{E}_{\underbrace{Pr_{i}|Pr_{0}=x_0}_{p_i}} \Big[  \prod_{i=1}^{N-1} \mathbb{E}_{h_i}[ \exp(-s h_i p_i] \Big] \\
& = \mathbb{E}_{p_i} \Big[  \prod_{i=1}^{N-1} \Big( 1 + \frac{s p_i}{m}\Big)^{-m} \Big]=  {\Big[  \mathbb{E}_{p_i} \Big[  \Big( 1 + \frac{s p_i}{m}\Big)^{-m} \Big] \Big]}^{N-1} \nonumber.
\end{align}
Now, applying the definition of mean and using the conditional PDF $f_{Pr_{i}|x_0}(p_i)$, Lemma 2 is obtained.

\section{Proof of Lemma 3}
Conditioned on $Pr_{0},Pr_I$, the received power from the $N-2$ interfering UAV-BSs $Pr_{i}$ are i.i.d. RVs and $Pr_I$ also determine the range of values for $Pr_{i}$. Accordingly, the set of $\{Pr_{i}\}_{i=2:N-1}$ now lies within $[0,Pr_I]$ with PDF of each element given by $f_{Pr_{i}|Pr_I}(p_i) = \frac{f_{Pr_{i}}(p_i)}{F_{Pr_{i}}(Pr_I)}$. Next, by the definition of mean and conditioned on the interference power from the dominant interfering UAV-BS is given by
\begin{equation}
\begin{split}
&\Omega(Pr_{0},Pr_I) = \mathbb{E}[I_{N-2}] \\
& = \mathbb{E}_{h_i,Pr_i}\Big[\sum_{i=2}^{N-2} h_i Pr_i|Pr_{0},Pr_I\Big] \\
& \myeqa (N-2) \mathbb{E}_{h_i,Pr_i}[ h_i Pr_i|Pr_{0},Pr_I] \\
& \myeqb (N-2) \mathbb{E}_{Pr_i}[ Pr_i|Pr_{0},Pr_I] \\
& \myeqc (N-2) \int_0^{Pr_I} p_i f_{Pr_{i}|Pr_I}(p_i)  {\rm{d}} p_i, \\
\end{split}
\end{equation}
where (a) follows the conditionally i.i.d. nature of $Pr_i$ conditioned on $Pr_{0}, Pr_I$ and independency between $Pr_i$ and $h_i$, (b) follows from $\mathbb{E}[h_i] = 1$ and (c) follows after applying the definition of mean using the PDF $f_{Pr_{i}|Pr_I}(p_i)$.

\section{Proof of Proposition 1}
Conditioned on $Pr_{0}=x_0$ and $Pr_{I}=x_I$, the conditional coverage probability under the dominant interferer approach is given by
\begin{equation}
\begin{split}
&\mathcal{P}_{c}(\theta | x_0, x_I)  \\
& = \mathbb{E}_{h_0,h_I}\Big[\mathbb{P}\Big[ h_0  > \frac{\theta h_I x_I + \theta \Omega(x_0,x_I)}{x_0} | x_0, x_I\Big] \Big] \\
& \myeqa  \mathbb{E}_{h}\Big[\mathbb{P}\Big[ \frac{\Gamma\big(m,m\frac{\theta h x_I + \theta \Omega(x_0,x_I) }{x_0}\big)}{\Gamma(m)} | x_0, x_I \Big] \Big] \\ 
& \myeqb\int_{0}^{\infty} \mathbb{P}\Big[ \frac{\Gamma\big(m, m\frac{\theta h x_I + \theta \Omega(x_0,x_I) }{x_0}\big)}{\Gamma(m)} | x_0, x_I \Big] f_{h_i}(h) {\rm{d}} h, 
\end{split}
\end{equation}
where (a) follows from the complementary CDF (cCDF) of the Gamma RV $h_0$, (b) follows after applying  the definition of mean for $h_I$. Finally, Proposition 1 results after deconditioning with respect to the joint PDF $f_{Pr_{0},Pr_{I}}(x_0,x_I)$  given by Lemma 4 for the RVs $Pr_{0}$ and $Pr_{I}$.

\section{Proof of Lemma 5}
By following a similar approach as in Appendix A, PDF $f_{Pr_{y}}(x)$ is obtained as in (22) as an intermediate step in calculating the PDF of maximum received power. The CDF $F_{Pr_{y}}(x)$ can then be derived as shown in Lemma 1.  
The PDF of $Pr_0 = \underset{y \in \Phi }{\operatorname{max}} \{Pr_{y}\} $ can now be obtained by exploiting the results of the Order Statistics \cite{orderstatistics}.  However, $Pr_0$ is meaningfully defined conditioned on $\mathcal{A}_{L(-R,R)}$. Accordingly, let $\mathcal{A}_{L(-R,R),n}$, with $n \geq 1$, define the event that $n$ UAV-BSs exist in $L(-R,R)$. Then, from the properties of the 1D $\Phi$, $\mathbb{P}[\mathcal{A}_{L(-R,R),n}] = e^{-\lambda_{UAV} |L(-R,R)|} (\lambda_{UAV} |L(-R,R)|)^n/n!$  Conditioned on $\mathcal{A}_{L(-R,R),n}$ and since the elements of $Pr_{y}$ are i.i.d., the probability that $Pr_0 \leq s_0$ is given by $\mathbb{P}[Pr_0 \leq s_0|\mathcal{A}_{L(-R,R),n}]=F_{Pr_0}(s_0|\mathcal{A}_{L(-R,R),n})= (F_{Pr_{y}}(s_0))^n$. Next, by exploiting the law of total probability, $F_{Pr_0|\mathcal{A}_{L(-R,R)}}(s_0)$ can be obtained as
\begin{equation}
F_{Pr_0|\mathcal{A}_{L(-R,R)}}(s_0) = \frac{\mathbb{P}[Pr_0 \leq s_0, \mathcal{A}_{L(-R,R)}]}{\mathbb{P}[\mathcal{A}_{L(-R,R)} ]}  .
\end{equation}
Now, $\mathbb{P}[Pr_0 \leq s_0, \mathcal{A}_{L(-R,R)}]$ can be derived summing $\mathbb{P}[Pr_0 \leq s_0, \mathcal{A}_{L(-R,R),n}]$ over all possible values of $n$, that is 
\begin{align}
& \mathbb{P}[Pr_0 \leq s_0, \mathcal{A}_{L(-R,R)}] =\sum_{n=1}^{\infty} (F_{Pr_{y}}(s_0))^n \mathcal{A}_{L(-R,R),n}\nonumber \\
& \myeqa \sum_{n=1}^{\infty} (F_{Pr_{y}}(s_0))^n e^{-\lambda_{UAV} |L(-R,R)|} \frac{(\lambda_{UAV}|L(-R,R)|)^n}{n!} \nonumber\\
& \myeqb e^{\lambda_{UAV} |L(-R,R)| \big(1-F_{Pr_{y}}(s_0)\big)} - e^{-\lambda_{UAV} |L(-R,R)| },
\end{align}
for $s_0 \in [0,\infty)$, where (a) follows after averaging over $n$ and (b) follows from $\sum_{n=1}^{\infty} \frac{a^n b^n}{n!} \myeqdef e^{a b} -1$. By substituting all the above in the definition of $F_{Pr_0|\mathcal{A}_{L(-R,R)}}(s_0)$, $F_{Pr_0|\mathcal{A}_{L(-R,R)}}(s_0)$ yields. Finally, the PDF $ f_{Pr_0|\mathcal{A}_{R_L,1}}(s_0)$ is obtained directly as $f_{Pr_0|\mathcal{A}_{R_L,1}}(s_0) = \frac{d F_{Pr_0| \mathcal{A}_{L(-R,R)}}(s_0)}{d s_0} = \frac{ \lambda_{UAV}|L(-R,R)| \,f_{Pr_{y}}(s_0) \,e^{\lambda_{UAV} |L(-R,R)| \big( F_{Pr_{y}}(s_0) - 1 \big) }}{1 - e^{- \lambda |L(-R,R)|}} $.

\section{Proof of Lemma 6}
The Laplace transform of $I$ conditioned on $s_0$  is given by \eqref{AppF1}, shown at the bottom of the page, where (a) follows from the  moment generating function (MGF) of $h_y$, (b) follows from the probability generating functional (PGFL) of the 1D PPP and considering the transformation $d_y=\sqrt{\|y\|^2+h^2}$. Then, the differential ${\rm d} \|y\| = \frac{{\rm{d}}\Big(\sqrt{d_y^2-h^2}\Big)}{{\rm{d}}d_y}{\rm{d}}d_y $. Moreover, in (b) we average over all values of $s_y$ w.r.t to the PDF $f_{S_y}(s_y)$. Finally, (c) follows from integration over the set of the region $\mathbf{\Omega}$, which has already been derived.

\begin{figure*}[!hb]
\hrulefill
\begin{equation} \label{AppF1}
\begin{split}
&\mathcal{L} _{I}(s|s_0)=\mathbb{E}_{\Phi^{!}}[e^{-sI}]=\mathbb{E}_{\Phi^{!}}\Big[e^{-s \sum_{y \in \Phi^{!}}  h_y S_{y} l(d_y)}\big|s_0\Big] \\
&\myeqa \mathbb{E}_{\Phi^{!}}\Big[\prod_{y \in \Phi^{!}} \Big(1+\frac{s\,  S_{y}\, l(d_y)  )}{m} \Big)^{-m} \big|s_0 \Big] \\
&\myeqb {\rm{exp}}\Bigg( -2\lambda_{UAV} \iint_\mathbf{\Omega} \Big(1- \Big(1+\frac{s\, s_y\, d_y^{-\alpha}}{m} \Big)^{-m}\Big) \frac{d_y}{\sqrt{d_y^2-h^2}}f_{S_y}(s_y) {\rm d} S_y {\rm d} d_y \Bigg) \\
&\myeqc {\rm{exp}}\Bigg(-2\lambda_{UAV} \int_{h}^{\sqrt{h^2+R^2}} \int_{0}^{s_0 d_y^{\alpha}} \Big(1-\Big(1+\frac{s\, s_y\, d_y^{-\alpha}}{m} \Big)^{-m}  \Big)  \frac{d_y}{\sqrt{d_y^2-h^2}}f_{S_y}(s_y) {\rm d} S_y {\rm d} d_y \Bigg)
\end{split}
\end{equation}
\end{figure*}

\bibliographystyle{IEEEtran}
\bibliography{IEEEabrv,references}
\balance

\end{document}